\documentclass[traditabstract]{aa}

\usepackage{natbib}

\def\ApJ{ Ap. J.}

\def\MNRAS{ Mon. Not. Roy. Astron. Soc.}

\def\PRL{ Phys. Rev. Lett.}
\def\PRD{ Phys. Rev. D}

\begin{document}
\title{Friedmann-free limits on spatial curvature}
\author{ James Rich\inst{1}    
}
\institute{CEA, Centre de Saclay, Irfu/SPP,  F-91191 Gif-sur-Yvette, France 
}
\date{Received xx; accepted xx}
\authorrunning{J. Rich}
\titlerunning{Friedmann-free limits on spatial curvature}
\abstract{
We discuss limits on cosmological spatial curvature
that can be derived without imposing the geometry-density relation
required by the Friedmann equation.  In particular,
studies of the expansion history using stellar evolution in
passive galaxies imply a curvature radius greater than the
Hubble distance.
}
\maketitle

The standard $\Lambda CDM$ cosmological model has been deduced from
astronomical data that are generally 
interpreted within the Friedmann-Lema\^itre-Robertson-Walker
framework for a homogeneous universe.
The framework uses the Robertson-Walker metric with the proper time
as a function of co-moving coordinates given by
\begin{equation}
d\tau^2 = dt^2 - a(t)^2\left[ \frac{dr^2}{1-Kr^2}
+ r^2(d\theta^2+\sin^2\theta d\phi^2) \right] \;.
\label{robertsonwalker}
\end{equation}
The dimensionless scale parameter $a(t)$ reflects the universal expansion
relative to the present epoch, $t_0$, $a(t_0)=1$.
The spatial curvature parameter, $K$, can be positive or
negative and is related to the curvature radius by $r_c=1/\sqrt{|K|}$.
The metric allows one to calculate trajectories of test particles
(e.g. photons) in terms of $a(t)$ and $K$ which are
governed by the Friedmann equation:
\begin{equation}
H(t)^2 \equiv \left(\frac{\dot{a}}{a}\right)^2 =
\frac{8\pi}{3} G\rho(t) - \frac{K}{a^2}
\label{friedmann}
\end{equation}
for an energy density $\rho(t)$.

Cosmological data analyzed in this framework imply
a nearly flat ($r_c \gg c/H_0$) 
model where the universal expansion is accelerating
($\ddot{a}>0$).
Both of these conclusions concern parameters of the Robertson-Walker
metric so
it is interesting to see to what extent they  can
be derived without assuming that $a(t)$ is governed by the
Friedmann equation, i.e., assuming homogeneity but not any particular
gravitational dynamics.
This is especially true given that the observed acceleration has
encouraged speculation that gravity might be modified at cosmological scales.

By measuring directly the deceleration parameter, $q_0=H_0^{-2}\ddot{a}/a$,
data on moderate-redshift type Ia supernovae  (SNIae)
imply acceleration independent of the Friedmann equation \citep{turner}.
\citet{sutherland}  has recently  proposed
a Friedmann-free test for
acceleration  using baryon acoustic oscillation (BAO) effects
in the matter correlation function.

Studying spatial curvature without using the Friedmann equation
is a bit trickier.
The very strong published limits on $r_c$ use the Friedmann equation
evaluated at the present epoch which yields
the geometry-density relation:
\begin{equation}
r_c = \frac{c/H_0}{\sqrt{k(\Omega_T -1)}}
\hspace*{10mm}k=\frac{K}{|K|}
\label{a0fromfried}
\end{equation}
where $H_0$ is the present value of the expansion rate and
where $\Omega_T$ is the present energy density in units
of the critical density $3H_0^2/8\pi G$.
Cosmic microwave background (CMB) data combined with BAO and SNIa data
imply $-0.0178<(1-\Omega_T)<0.0063$ \citep{wmap}
which gives
\begin{displaymath}
r_c > 7.5(c/H_0) \hspace*{5mm}K<0\;\; ,
\end{displaymath}
\begin{equation}
r_c > 12.5(c/H_0) \hspace*{5mm}K>0\;\; .
\label{limitwithfried}
\end{equation}

In the cosmology of a homogeneous universe, the spatial curvature
determines the relation between the  present distance, $d(z)$, to an object
of redshift $z$ and the luminosity and angular distances, $d_L(z)$
and $d_A(z)$.
These two distances determine the flux, $F$, from objects
of luminosity, $L$, and the angular size, $\Delta\theta$, of an object
of size $R$:
\begin{displaymath}
F = \frac{L}{4\pi d_L^2} \hspace*{10mm} \Delta\theta=\frac{R}{d_A} \;\;.
\end{displaymath}
In the absence of curvature, $d_L$ and $d_A$ are both proportional to
$d(z)$, with factors of proportionality $(1+z)$ for $d_L$ and $(1+z)^{-1}$ for
$d_A$.
For universes with $K>0$ ($K<0$),
$d_L$ and $d_A$
are less than (greater than) what would be expected from proportionality.

It is not easy to test this fundamental prediction.
The basic problem is that while $d_A$ and $d_L$ can be determined
directly by using standard candles and rulers,
$d(z)$ can only be calculated using knowledge of the expansion rate:
\begin{equation}
d(z)
= \int_{t_1}^{t_0} \frac{dt}{a(t)}
= \int_{a_1}^{1} \frac{da}{a \dot{a}}
= \int_0^z \frac{dz}{H(z)}
\label{lightcone}
\end{equation}
where 
the redshift is given by $z(t)+1=1/a(t)$.
The limits of integration are given by the times of signal emission, $t_1$,
and reception, $t_0$.
The luminosity and angular distances depend on $r_c$:
\begin{equation}
d_{L,A}(z) = (1+z)^{(1,-1)}r_c S_k(d(z)/r_c)
\label{dldrelation}
\end{equation}
where  $S_1(x)=\sin(x)$ and $S_{-1}(x)=\sinh(x)$ for 
$K>0$ and $K<0$.
Equation (\ref{dldrelation})
defines the relation between $d(z)$ and $d_{L,A}(z)$ via
the curvature radius $r_c$.
If $d(z)$ were known, it would allow a determination of $r_c$
and a verification that it takes the 
value (\ref{a0fromfried}).

If $a(t)$  and/or $H(t)$ are not directly measured,
$d(z)$ can only be calculated by using
the Friedmann equation in equation (\ref{lightcone}).
This imposes
the value of $r_c$ given by (\ref{a0fromfried}).
As such, the fundamental relationship (\ref{dldrelation}) cannot be tested
in this way.  Rather, the luminosity and angular distances are
used to determine cosmological parameters
($\Omega_M,\Omega_\Lambda,\Omega_T$) that can then be used to
calculate the distance $d(z)$.

To avoid use of the Friedmann equation, one needs an independent
measurement of $H(z)$ to be injected into equation ({\ref{lightcone}).
Probably the best possibility is to use the radial BAO feature
in the matter correlation function which gives directly $H(z)$.
\citet{clarkson} 
noted that this
could directly measure spatial curvature.
Note that while the length of the BAO standard ruler depends on
the Friedmann equation in the pre-recombination universe, the measurement
of $H(z)$ depends only 
on it being a co-moving ruler, i.e. that
galaxies having nearly fixed co-moving coordinates.
This is a non-trivial assumption but one that does not depend explicitly
on the Friedmann equation.

In the absence of  BAO measurements of $H(z)$, we note that
if galaxies were equipped with clocks that we could see
and read, measuring distances
would be trivial since an ensemble of such galaxies
at different redshifts would yield $z(t)$  from
redshift and time measurements 
\citep{jimenez}.
Distances could then be calculated using equation (\ref{lightcone})
without imposing the Friedmann equation.

Stellar evolution in  passive
galaxies (those with negligible star formation)  can be used as such a
clock.
The first steps in their use  for cosmology
have been performed by \citet{simon}, 
\citet{figueroa} 
and 
\citet{moresco}.
They derive expansion histories that are consistent
with the standard flat $\Lambda CDM$ cosmology
($\Omega_\Lambda=0.73$ and $\Omega_M=0.27$).
The most precise results were obtained by 
\citet{moresco} who observed the redshift
evolution of the  4000\AA-spectral break in SDSS elliptical
galaxies in the range $0.15<z<0.30$.
Stellar population evolution models then gave the elapsed
time between $z=0.15$ and $z=0.3$ necessary for the
spectral evolution.
Using $H(z)=\dot{z}/(1+z)$,
this gives the expansion rate for $0.15<z<0.30$.
They extrapolated this rate to zero redshift 
using the standard cosmology
 and found a value
$H_0=72.6\pm2.9(stat)\pm2.3(syst){\rm km\,s^{-1}Mpc^{-1}}$,
that is in agreement with recent local measurements,
$H_0=(74.2\pm3.6){\rm km\,s^{-1}Mpc^{-1}}$ 
\citep{hzero}.
From our point of view,
this means that at the 5\% level
the  expansion rate in the range $0<z<0.3$ follows 
the flat $\Lambda CDM$ prediction.

Since the measured expansion rate follows the flat $\Lambda CDM$
prediction,
equation (\ref{lightcone}) implies that the distances
$d(z)$ to objects in this redshift range agree 
with the flat $\Lambda CDM$ values.
Furthermore,
the  mean luminosity distances in this redshift range 
from SNIa
\citep{sn1a}
agree with those of the standard flat $\Lambda CDM$ model to
a precision of better than 5\%.
The agreement of both $d(z)$ and $d_L(z)$ with the same flat model
implies that  spatial curvature effects must be small at $z=0.3$.

To quantify the limit on $r_c$, we
invert (\ref{dldrelation}),
giving $d(z)$ as a function of $d_L(z)$, and then take the derivative
with respect to $z$.  Using $d^\prime(z)=1/H(z)$ and the fact
that $d_L(z)$ is well described by a flat $\Lambda CDM$ model,
one finds
\begin{equation}
\frac{H(z)}{H_{flat}(z)} = 
\sqrt{1
-k\left(
\frac{d_{flat}(z)}{r_c}
\right)^2
}
\hspace*{10mm}k=\frac{K}{|K|}
 \;\;,
\end{equation}
where $H_{flat}$ and $d_{flat}$ are the expansion rate and distance
for the flat $\Lambda CDM$ model that gives the observed luminosity
distance.
For the measurement of $H(z\sim0.22)$ of \citet{moresco}, if we adopt
a conservative limit 
\begin{equation}
0.8<\frac{H(z=0.22)}{H_{flat}(z=0.22)}<1.2
\end{equation}
we find
\begin{equation}
r_c > 1.5d_{flat}(0.22) \sim  0.3(c/H_0) \hspace*{5mm}K<0\;\; ,
\end{equation}
\begin{equation}
r_c > 1.6d_{flat}(0.22) \sim  0.3(c/H_0) \hspace*{5mm}K>0\;\; .
\end{equation}

The work of 
\citet{figueroa} 
provides $H(z)$ up to $z\sim1.5$, but at a precision
of only $\sim20\%$.
Adopting
\begin{equation}
0.6<\frac{H(z=1.5)}{H_{flat}(z=1.5)}<1.4
\end{equation}
we find
\begin{equation}
r_c > d_{flat}(1.5) \sim  (c/H_0) 
\hspace*{5mm}K<0
\;\; ,
\end{equation}
\begin{equation}
r_c > 1.25d_{flat}(1.5) \sim  1.3(c/H_0)
 \hspace*{5mm}K>0
\;\; .
\end{equation}
Here again, we use the fact that 
SNIa for $0.3<z<1.5$ \citep{guy,hstsn} give
luminosity distances that are in agreement with flat $\Lambda CDM$.
Baryon acoustic oscillations measurements 
\citep{bao,percival,wigglez} imply the same  
for the distance combination
$d_A(z)^2c/H(z)$ ($0.2<z<0.6$ ).

The limits on $r_c$ obtained here are rather modest, implying
only $\Omega_T<1.6$ if one assumes the validity of the 
Friedmann equation.
It will be possible to confirm and improve the limits in the future
with BAO $H(z)$ measurements \citep{boss,anze}.  We note that BAO has
the advantage of  not depending on stellar modeling.
A particularly interesting challenge would be to perform
a Friedmann-free CMB analysis with the hope of extending
the limits to $z=1070$.
Unfortunately, unless our ideas of gravity are wrong at
cosmological scales and the Friedmann equation does not apply,
the present limits on $\Omega_T$ mean that it will be
very difficult to directly see cosmological curvature
with these methods.

\begin{acknowledgements}
I thank
Raul Jimenez, Jean-Baptiste Melin, Will Sutherland and Jean-Philippe Uzan
for stimulating discussions.
I thank the referee for comments that improved the paper.
\end{acknowledgements}

\end{document}